\documentclass[twocolumn,amssymb,amsmath,nobibnotes,nofootinbib,showpacs,aps, prb]{revtex4}
\usepackage[dvips]{graphicx}
\begin{document}
\title{Magnetic and magnetoresistive properties of cubic Laves phase HoAl$_2$ single crystal}
\author{M. Patra}
\affiliation{Department of Physics, Raghunathpur College, Raghunathpur, Purulia 723133, India}
\author{S. Majumdar}
\author{S. Giri}
\affiliation{Department of Solid State Physics, Indian Association for the Cultivation of Science, Jadavpur, Kolkata 700 032, India}
\author{Y. Xiao} 
\affiliation{Forschungszentrum Juelich, D52425 Juelich, Germany}
\author{T. Chatterji}
\affiliation{Institut Laue-Langevin, BP 156, 38042 Grenoble Cedex 9, France}
\begin{abstract}
We report the magnetization ($M$) and magnetoresistance (MR) results of HoAl$_2$ single crystals oriented along $<100>$ and $<110>$ directions. Although HoAl$_2$ has cubic Laves phase structure, a large anisotropy is observed in $M$ and MR results below Curie temperature ($T_C$) at 31.5 K. A satisfactory correlation between magnetic entropy change ($\Delta S_M$) and MR could be established along $<110>$ direction and in $<100>$ direction except for the temperature ($T$) region, around which spin reorientation takes place. Large inverse magnetocaloric effect is observed at low-$T$ which is associated with the spin reorientation process in $<100>$ direction. A theoretical model based on Landau theory of phase transition can describe $T$-variation of -$\Delta S_M$ for $T > T_C$. 

\end{abstract}

\pacs{71.20.Lp, 75.30-Sg, 72.15.-v}

\maketitle

\section{Introduction}
The rare earth intermetallic compounds, $R$Al$_2$ ($R$ = rare earth) draw significant attention for intriguing thermoelectric, \cite{sakurai} magnetoelastic, \cite{ibarra1} domain dynamics. \cite{levin} In this series of compounds only $R$ is magnetic whereas Al is magnetically neutral. Thus, long range magnetic order is settled by indirect spin exchange interaction between localized magnetic moments of $R$-atoms via conduction electrons. \cite{purw} Magnetic properties of $R$Al$_2$ series are found to be dominated by the delicate interplay between exchange interaction and crystalline electric field (CEF) effect. The considerable CEF typically removes the degeneracy of multiplate ground state, leading to the anomalous effect in magnetic properties. 

The series of compounds, $R$Al$_2$ are fascinating for large magnetocaloric entropy change due to change in external magnetic field. \cite{ranke1,ranke2,ranke3,campoy,ranke4,lima} This large change in magnetocaloric effect (MCE) is  attractive for the technological applications in magnetic refrigeration due to its potential advantage of environmental friendliness over gas refrigeration. Current research is strongly motivated on the microscopic origins of large  MCE that are accountable for improving magnetic refrigeration. \cite{gasch} Since $R$Al$_2$ series is promising for large MCE, various fundamental aspects of MCE in $R$Al$_2$ ($R$ = rare earth) has been probed experimentally and theoretically in the last decades. \cite{ranke1,ranke2,ranke3,campoy,ranke4,lima} Recently, a nice correlation between magnetoresistance (MR) and MCE has been demonstrated in $R$Al$_2$ ($R$ = Pr, Nd, Tb, Dy, Ho, and Er). \cite{campoy} Anomalous MCE  was experimentally observed along $<111>$ direction of DyAl$_2$ which has been successfully explained by a mean field model considering interplay between spin reorientation, exchange interaction, and CEF effects. \cite{lima} Extensive theoretical investigations have been performed in this series of compounds where magnetic exchange interaction and CEF effect were taken into account to interpret experimentally observed MCE results. \cite{ranke1,ranke2,ranke3,ranke4} 

Rare-earth intermetallic compound, HoAl$_2$ crystallizes in cubic Laves phase (C15) and  orders ferromagnetically at the Curie temperature, $T_C$ = 31.5 K.\cite{schelp} Thermal variation of heat capacity exhibits a pronounced signature at $T_C$. \cite{hill} In addition to the  appearance of a peak at $T_C$, another well defined peak was noticed at 20 K. This peak was defined as spin reorientation temperature ($T_r$) and this appeared when external magnetic field was non-coincident with the easy-axis direction. Appearance of spin reorientation was further confirmed through magnetoelastic behavior \cite{ibarra1} and torque magnetometry. \cite{ibarra2} Recently, a theoretical investigation of MCE was performed in detail on HoAl$_2$ where a valley was predicted in thermal variation of MCE. \cite{oliveira} This valley was suggested due to spin reorientation at low temperature from $<110>$ easy direction toward applied field direction along $<100>$. 

In this article, we present magnetocaloric properties of single crystalline HoAl$_2$ both from the temperature and field dependent magnetization and MR results along $<100>$ and $<110>$ directions. A large anisotropy in magnetic and magnetoresistance results along major crystallographic axes is evidenced below $T_C$, although HoAl$_2$ has cubic Laves phase structure. The clear signature of spin reorientation is noticed both in magnetization and MR results in $<100>$ direction which was not so apparent in the polycrystalline results. \cite{campoy} Analogous to that observed in polycrystalline compound \cite{campoy} a fair agreement between magnetization and MR results has been established in $<110>$ direction.  Interestingly, low temperature results in $<100>$ direction, around which spin reorientation process takes place, does not correlate each other and displays different results than that of the theoretical prediction. \cite{oliveira}

\section{Experimental details}
A large crystal of HoAl$_2$ was cut in to two smaller crystals such that magnetic field can be applied parallel to $<100>$ and $<110>$ crystallographic directions. Magnetization was measured in a commercial superconducting quantum interference device (SQUID) magnetometer (MPMS) in the temperature range from 2 to 300 K and magnetic field up to 7 T. The resistivity ($\rho$) was measured by using a commercial Physical Property Measurement System (PPMS) of Quantum Design. A standard four-probe method was used for electrical resistance measurements where silver epoxy was used for electrical contacts. The MR measurements were carried out at constant temperature by applying magnetic field ($H$) from 0 to 70 kOe and at fixed $H$ by varying temperature in the range 4 - 100 K. MR is defined as [$\rho(H) - \rho(0)]/\rho(0)$, where $\rho(H)$ and $\rho(0)$ represent $\rho$ in a static field and zero field, respectively.

To calculate MCE isothermal magnetization curves ($M-H$) were measured from 4 to 100 K by changing magnetic field from 0 to 70 kOe. The change in magnetic entropy i.e MCE was  calculated using the following equation derived from Maxwell's thermodynamic relation as
\begin{equation} 
\Delta S_M = \int_0^{H} \left(\frac{\partial M}{\partial T}\right)_H dH.
\end{equation}

\begin{figure}[t]
\vskip 0.4 cm
\centering
\includegraphics[width = \columnwidth]{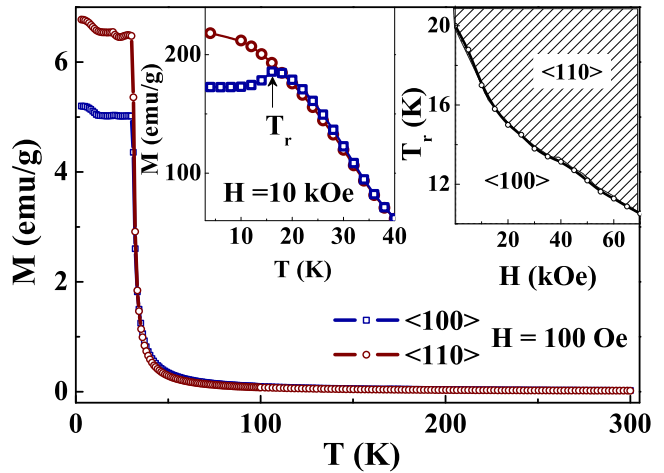}
\caption {Thermal variation of magnetization $(M-T)$ for applying field ($H$ = 100 Oe) parallel to $<100>$ and $<110>$ directions. Left inset shows the $M-T$ curves at $H$ = 10 kOe and right inset shows the spin re-orientation temperature ($T_r$) - $H$ phase diagram.}
\label{Fig. 1}
\end{figure}
\section{Experimental results and discussions}

\subsection{Magnetization}
Thermal variation of the magnetization ($M$) along easy axis $<110>$ and $<100>$ directions measured at $H$ = 100 Oe is displayed in Fig. 1. This is in accordance with the magnetization results reported in polycrystalline HoAl$_2$ \cite{campoy} and other experimental studies \cite{schelp,hill} as well as theoretical calculations \cite{oliveira} where the compound orders ferromagnetically at 31.5 K ($T_C$). Paramagnetic region could be fitted in Curie-Weiss behavior with effective paramagnetic moment, $\mu_{\rm eff}$ = 10.48 $\mu_{\rm B}$ and Curie-Weiss temperature, $\Theta_{\rm CW}$ = 33.5 K. The $\Theta_{\rm CW}$ is slightly larger than $T_C$. The value of $\mu_{\rm eff}$ is close to the theoretical value (10.6 $\mu_{\rm B}$) of Ho$^{3+}$. We measured thermal hysteresis close to $T_C$ which is absent in the current investigation, suggesting a second order paramagnetic to ferromagnetic (FM) phase transition. As seen in Fig. 1 the value of magnetization in easy direction is larger than the other direction along $<100>$. This is in accordance with the high-field results where values of magnetization at 4 K and 70 kOe are 8.92 $\mu_{\rm B}$ and 8.56 $\mu_{\rm B}$ in $<110>$ and $<100>$ directions, respectively. We note that magnetization at 4 K saturates in easy direction, although it does not saturate even at 70 kOe in $<100>$ direction. When the magnetic field of 10 kOe was applied parallel to $<100>$, a peak in $M-T$ curve appears at 17 K as seen in the left inset of Fig. 1. This peak appears due to spin reorientation and strongly depends on magnitude of $H$. As seen in the $T_r - H$ phase diagram given in the right inset of Fig. 1, $T_r$ decreases with increasing $H$ from 20 K to 10.5 K for $H$ = 0.1 kOe to 70 kOe. The value of $T_r$ at 20 K measured at 0.1 kOe exactly matches with the reported values of $T_r$ found in the literatures from different measurement techniques.  \cite{hill,ibarra1,ibarra2,campoy} 
\begin{figure}[t]
\vskip 0.4 cm
\centering
\includegraphics[width = \columnwidth]{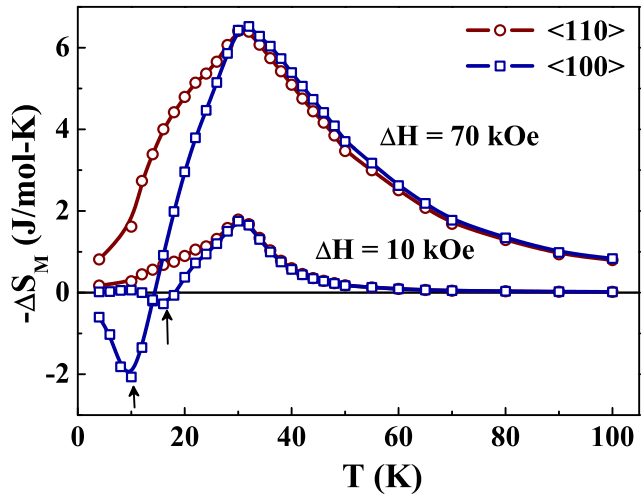}
\caption {Thermal variation of magnetic entropy change (-$\Delta S_M$) for field parallel to $<100>$ and $<110>$ directions with change in field ($\Delta H$) = 10 and 70 kOe.}
\label{Fig. 2}
\end{figure} 
\par    
The magnetic entropy change, -$\Delta S_M$ is calculated from isothermal magnetization curves ($M-H$) using Eq. (1). Thermal variations of -$\Delta S_M$ for $\Delta H$ = 10 and 70 kOe in the range 4 - 100 K are displayed in Fig. 2 where field was applied parallel to $<110>$ and $<100>$ directions. As expected a large peak is noticed at $T_C$. At $T_C$ the value of -$\Delta S_M$ is  $\approx$ 6.5 J/mol-K which is larger than that observed in the  polycrystalline compound (5.5 J/mol-K). \cite{campoy} The components along $<110>$ and $<100>$ directions deviate from each other below $T_C$. The -$\Delta S_M - T$ plot in $<110>$ direction is analogous to that observed in polycrystalline compound. \cite{campoy} Interestingly, a new feature is observed at low temperature when field was applied parallel to $<100>$ direction. Inverse MCE or change of sign in -$\Delta S_M - T$ plot is noticed at low temperature. At further lower temperature a deep is observed in the -$\Delta S_M-T$ plot. The temperature, at which deep is observed, varies with $\Delta H$. We note that this deep occurs at same temperature, where signature of $T_r$ is noticed in the $M - T$ curve. Signature of $T_r$ in -$\Delta S_M - T$ plot was already predicted by Oliveira {\it et al}. \cite{oliveira} from the theoretical calculations. Current investigation confirms evidence of spin reorientation in $<100>$ direction. We note that nature of -$\Delta S_M - T$ curve around $T_r$ is dissimilar to the theoretical calculations when -$\Delta S_M - T$ is plotted at different $\Delta H$. The deep in the -$\Delta S_M - T$ plot at $T_r$ is pushed downward with increasing $\Delta H$. The value of positive magnetic entropy change is increased to 2.07 J/mol-K from 0.28 J/mol-K for $\Delta H$ increased to 70 kOe from 10 kOe. These results are analogous to that observed in single crystalline, DyAl$_2$ in $<111>$ direction. \cite{lima} The emergence of inverse MCE attracts the community for probing the underline mechanism in magnetism as found in the literatures. \cite{chat1,manosa1,patra1} Inverse MCE was reported in shape memory alloys which has been involved with the martensitic phase transformation. \cite{chat1,manosa1} Inverse MCE involved with the FM/antiferromagnetic (AFM) interactions has also been proposed in manganites \cite{patra1} and in various compounds reviewed by von Ranke {\it et al}. \cite{ranke5} Here, inverse MCE is correlated to the decrease in $M$ below $T_r$ in $<100>$ direction. This primarily occurs due to rotation of magnetization from easy direction to $<100>$. 
\begin{figure}[t]
\vskip 0.4 cm
\centering
\includegraphics[width = \columnwidth]{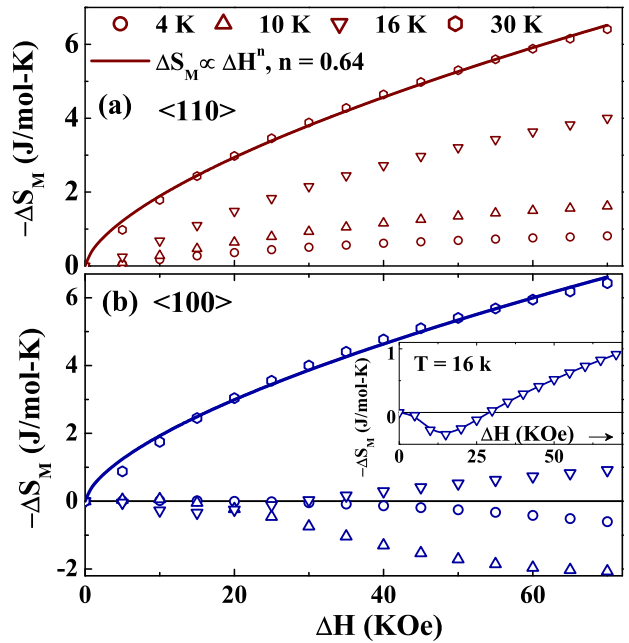}
\caption {Magnetic field ($H$) dependence of -$\Delta S_M$ at different temperatures along $<110>$ and $<100>$ directions.}
\label{Fig. 3}
\end{figure} 
\par
Figure 3 depicts magnetic field dependence of -$\Delta S_M$ at different representative  temperatures in $<110>$ and $<100>$ directions. According to mean-field approximation -$\Delta S_M$ can be expressed as -$\Delta S_M \propto H^n$, where the exponent, $n$ = 2/3 at $T_C$. \cite{oest} We note that this mean-field approach satisfies field dependence of -$\Delta S_M$ close to $T_C$ with $n$ = 0.64. This is displayed in Figs. 3(a) and 3(b) where continuous curves show the fits at 30 K in both the $<110>$ and $<100>$ directions. The figure further displays the curves at representative temperatures at 4 and 10 K (below $T_r$) and at 16 K, which lies in the range of $T_r$, 10.5 - 20 K for the measurement of $\Delta H$ in between 0.1 - 70 kOe. As expected negative MCE is seen for all the temperatures and $\Delta H$ in between 10 - 70 kOe for the measurement in $<110>$ direction. On the other hand, positive MCE is noted at 4 and 10 K which are below $T_r$. Negative MCE is observed at 30 K (well above $T_r$). The curve at 16 K exhibits anomalous behavior where it initially shows small positive MCE and then moves to negative MCE for $\Delta H >$ 30 kOe. This is elaborated in the inset of Fig. 3(b). We note that maximum positive MCE is observed at 15 kOe, at which $T_r$ is observed at 16 K. Analogous -$\Delta S_M-H$ curves have been reported in the literatures where anomalous behavior was correlated to the coexistence of FM and AFM interactions. \cite{patra1} Here, this anomalous behavior is involved with the spin reorientation processes while magnetic field is applied in $<100>$ direction.  

\begin{figure}[t]
\vskip 0.4 cm
\centering
\includegraphics[width = \columnwidth]{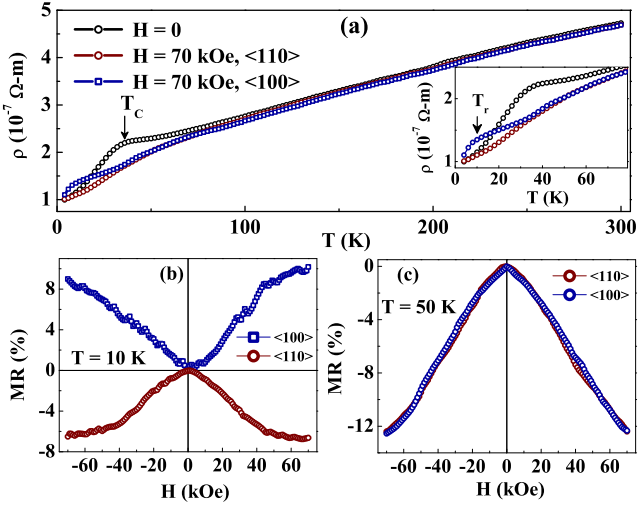}
\caption {(a) Temperature variation of $\rho(H,T)$ in $<100>$ and $<110>$ directions. Inset highlights the region below $T_C$. (b) and (c) show MR with $H$ in $<100>$ and $<110>$ directions at 10 K and 50 K, respectively.}
\label{Fig. 4}
\end{figure}

\subsection{Resistivity}
Temperature and magnetic field variation in $<100>$ and $<110>$ directions of $\rho(T,H)$ are carried out and the results are summarized in Fig. 4. In Fig. 4(a) temperature dependence of $\rho(0,T)$ is displayed. At high temperature nearly linear behavior is noted which is associated to the nonmagnetic electron-phonon interaction. A broad maximum is observed at $T_C$ in $\rho(0,T)$ which is suggested due to magnetic scattering close to paramagnetic to FM transition. When magnetic field is applied ($H$ = 70 kOe), the maximum is smeared out due to considerable magnetic entropy change as seen in the figure. In accordance with the magnetization results $\rho(H,T)$ strongly deviates from each other below $T_C$. The negative MR in $<110>$ direction is observed over a wide temperature range, 4 - 100 K which is maximum at $T_C$. The scaled value of MR as a function of temperature is depicted in Fig. 5 which is described below in detail. Although almost same value of negative MR is observed in $<100>$ direction for $T \geq T_C$, it changes the sign and positive MR is noticed at low temperature. The value of positive MR is maximum close to $T_r$ which is highlighted in the inset of Fig. 4(a). Examples of MR-$H$ curves below (at 10 K) and above (50 K) $T_C$ are shown in Figs. 4(b) and 4(c), respectively. At 50 K negative MR-$H$ curve overlaps each other for both the directions. In accordance with the MR-$T$ results below $T_C$ a contrast feature in the MR-$H$ curves is observed at 10 K where MR is positive in $<100>$ and negative in $<110>$ directions. The results are also in line with the magnetic entropy change below $T_C$ involving spin reorientation process. 

\begin{figure}[t]
\vskip 0.4 cm
\centering
\includegraphics[width = \columnwidth]{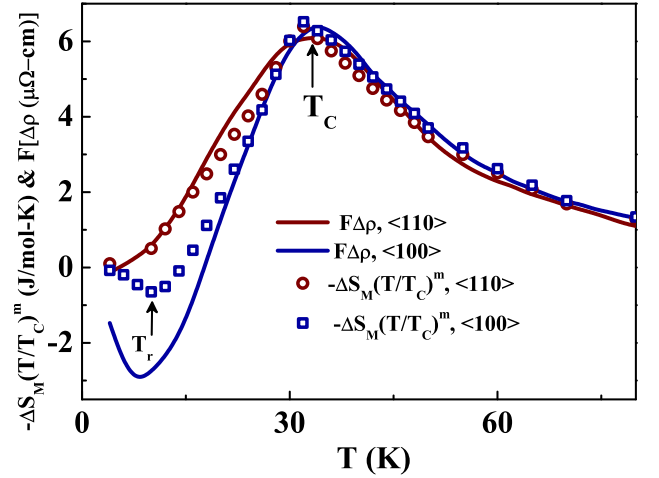}
\caption {Plots of $F \Delta \rho_M$ and -$\Delta S_M$(T/T$_C$)$^m$ vs $T$ curves for  $\Delta H$ = 70 kOe in $<100>$ and $<110>$ directions.}
\label{Fig. 5}
\end{figure} 
\par
Campoy {\it et al}. nicely demonstrated a correlation between -$\Delta S_M(T/T_C)^m$ and $F \Delta \rho_M$ for a series of polycrystalline intermetallic compounds, $R$Al$_2$. \cite{campoy} This correlation was also found to be satisfactory in hole doped manganites, exhibiting metal-insulator transition at $T_C$. \cite{xiong} Here, $\Delta \rho_M$ is defined as $\rho_M(H,T)-\rho_M(0,T) \approx \rho(H,T)-\rho(0,T) = \Delta \rho$ where $\rho_M(H,T)$ and $\rho_M(0,T)$ are the magnetic contribution to the resistivity in a static field and zero field, respectively. The above approximation is valid considering very small magnetoelastic effect and negligible change in phonon scattering, mobility and concentration of carriers due to application of magnetic field. $F$ is a factor which is determined to coincide maximum in both the -$\Delta S_M(T/T_C)^m-T$ and $F \Delta \rho_M-T$ curves at $T_C$. The value of $m$ is 1 for $T<T_C$ and $m$ = 0 for $T>T_C$. Plots of $F \Delta \rho_M-T$ and -$\Delta S_M(T/T_C$)$^m-T$ curves for $\Delta H$ = 70 kOe in $<100>$ and $<110>$ directions are shown in Fig. 5. Symbols are the $\Delta S_M(T/T_C)^m$ values as a function of temperature as obtained from the analysis of isotherm $M-H$ curves and continuous curves provide the scaled $F \Delta\rho_M$ with temperature. The values of $F$ are 1.2 and 1.4 in $<110>$ and $<100>$ directions, respectively. We note that the value along easy axis exactly matches with the calculation given in the literature. \cite{campoy} A nice agreement between these plots is noticed for $T \geq T_C$ in both the directions. In fact, a reasonable agreement is also observed below $T_C$ in $<110>$ direction. On the other hand, a considerable deviation is noted below 22 K in $<100>$ direction. These single crystal results clearly demonstrate that this correlation does not hold while spin reorientation process is involved.   

\subsection{Landau theory}
According to the Landau theory of phase transitions, Gibbs free energy can be expressed as
\begin{equation}
G(T,M) = G_0 + \frac{1}{2}AM^2 + \frac{1}{4}BM^4 + \frac{1}{2}CM^6 - MH,
\end{equation}
where the coefficients $A, B$, and $C$ are temperature dependent parameters usually known as Landau coefficients. The sign of $B$ is important, as it indicates the first-order transition
($B <$ 0) or the second-order transition ($B >$ 0). For energy minimization in Eq. (2) the equation of state is given by 
\begin{equation} 
H/M = A + BM^2 + CM^4.
\end{equation}
These Landau coefficients $A, B$, and $C$ are estimated from fits of the Arrott plots to Eq. (3) which are displayed by the solid lines in Fig. 6(a). The parameters, $A, B$, and $C$ obtained are plotted with $T$ in Figs. 6(b) and 6(c) for the measurement in $<110>$ direction. The parameter, $A$ standing for the inverse susceptibility changes the sign at $T_C$ as seen in Fig. 6(b). The positive values of parameters, $B$ over measured temperature range confirms the second order phase transition at $T_C$. The magnetic entropy change can be obtained by differentiating of Gibbs free energy with respect to temperature as
\begin{equation} 
\Delta S_M = -\frac{1}{2}A^{\prime}(T)M^2-\frac{1}{4}B^{\prime}(T)M^4-\frac{1}{6}C^{\prime}(T)M^6. 
\end{equation}
Thus, $\Delta S_M$ can be calculated using Eq. (4) where $A^{\prime}(T), B^{\prime}(T)$, and $C^{\prime}(T)$ are the temperature derivatives of Landau coefficients. We note that  coefficient of $M^6$ in (4) was not typically required to satisfy the experimental data  using this model even below $T_C$ for hole doped manganites. \cite{amaral} Here, this term is found to be mandatory. The calculated values of $\Delta S_M$ with $T$ around $T_C$ are depicted with the experimental data in $<100>$ and $<110>$ directions for $\Delta H$ = 70 kOe as seen in Fig. 6(d) by the continuous curves. The plot exactly matches the experimental data above $T_C$. However, this mean-field approximation can not interpret the results below $T_C$. The results are analogous to that observed in hole doped manganites where Landau theory nicely corroborated the experimental data for polycrystalline compound \cite{amaral} and can not satisfy the anisotropic MCE data of single crystal below $T_C$, although it could satisfy exactly for $T \geq T_C$. \cite{patra2} 
\begin{figure}[t]
\vskip 0.0 cm
\centering
\includegraphics[width = \columnwidth]{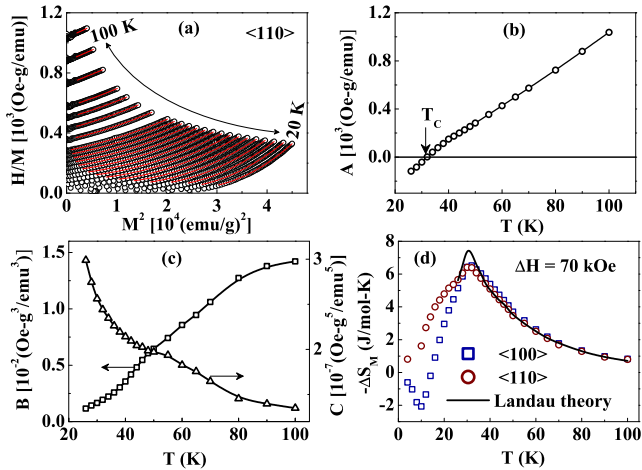}
\caption {(a) Arrott plot for applying field in $<110>$ direction. Solid lines are the fits using Eq. (3). (b) and (c) depicts temperature variation of $A$, $B$, and $C$. (d) Temperature variation of -$\Delta S_M$ and the fit (continuous curve) using Eq. (4).}
\label{Fig. 6}
\end{figure}
Recently, Oliveira {\it et al}. calculated $\Delta S_M$ by varying temperature at different magnetic field using a model Hamiltonian considering exchange magnetic interaction and  crystalline electric field anisotropy for HoAl$_2$. \cite{oliveira} Current investigation indicates that the calculation satisfies the single crystal data even below $T_C$, except for the low  temperature region, around which spin reorientation takes place. We note that slight modification of CEF parameters and adjustment of applied magnetic field in any crystallographic direction along with the inclusion of self consistency of free energy in the calculation can describe well the experimental $\Delta S_M$ value at low-$T$ involving spin reorientation along $<111>$ direction in DyAl$_2$. \cite{lima} This theoretical treatment might be useful to describe the low-$T$ MCE results in $<100>$ direction for HoAl$_2$. 

In the last decades, experimental investigations on MCE have been carried out in $R$Al$_2$ series of intermetallic compounds where most of them were studied in polycrystalline compounds. \cite{ranke1,ranke2,ranke3,campoy,ranke4,lima} A handful number of experimental studies have been performed on MCE in single crystals. \cite{lima} In few cases extensive theoretical calculations of magnetic entropy change adequately describe the polycrystalline data, which can not reproduce satisfactorily the anisotropic single crystal data below $T_C$. Therefore, single crystal results are very much desirable to identify the precise mechanism as well as proper theoretical understandings. The current magnetic and magnetoresistive results of single crystalline, HoAl$_2$ are fascinating, because this study can establish the anisotropy along different crytallographic directions. Furthermore, appearance of anisotropy in cubic HoAl$_2$ is interesting, which is suggested to be involved with the localized Ho$^{3+}$ ions having 4$f$ unpaired spins in presence of strong crystalline electric field. 

\section{Summary and Conclusions}
We measure magnetization and magnetoresistance of HoAl$_2$ single crystals in $<110>$ and $<100>$ directions. A considerable anisotropy is observed in $T$-dependent magnetization and MR results below $T_C$ in this cubic Laves phase compound which is suggested due to nonspherical 4$f$ wave function of localized Ho$^{3+}$ ions in crystalline electric field. A satisfactory correlation between $\Delta S_M$ obtained from isothermal magnetization curves and MR has been established in the measured $T$-range along $<110>$ and above $T_C$ in $<100>$ direction. The change of sign in $\Delta S_M$ or inverse MCE is observed in $<100>$ direction at low temperature which is suggested to be involved with spin reorientation process. Although a considerable negative MR is observed at $T_C$, analogous to that observed in $\Delta S_M$, MR changes the sign and convincingly a positive MR is noticed around spin reorientation temperature. The anomalous $T$- and $H$-dependent magnetization, MR, and $\Delta S_M$ in $<100>$ direction is identified at low temperature due to spin reorientation process which could be clearly evidenced in the single crystal data. 

An attempt is made to describe $T$-dependent $\Delta S_M$ theoretically along both the directions based on Landau theory of phase transition which can interpret a limited region  for $T > T_C$. Recently, Oliveira {\it et al}. \cite{oliveira} successfully explained  $T$-dependent $\Delta S_M$ in $<110>$ direction and a limited $T$-region in $<100>$ direction for HoAl$_2$. This can not describe the single crystal data in $<100>$ direction at low-$T$, around which spin reorientation process takes place. Lima {\it et al}. successfully described the $\Delta S_M-T$ curve involving spin reorientation process in DyAl$_2$ with slight modification of CEF parameters and magnetic interaction proposed by Oliviera {\it et al}. in their calculation. We suggest that this theoretical calculation carried out by Lima {\it et al}. may interpret the low-$T \Delta S_M$, around which spin reorientation takes place.

\end{document}